\documentclass[aps,prl,twocolumn,superscriptaddress,showpacs,
amsmath,amssymb,floatfix]{revtex4} 
\usepackage{graphicx} 
 
\newcommand{\beq}{\begin{equation}}   
\newcommand{\eeq}{\end{equation}}   
\newcommand{\bea}{\begin{eqnarray}}   
\newcommand{\eea}{\end{eqnarray}}   
   
\begin{document}      
\title{Dynamics of short time--scale energy relaxation of optical excitations 
due to electron--electron scattering in the presence of arbitrary disorder} 
\author{Imre Varga} 
\affiliation{Fachbereich Physik und Wissenschaftliches Zentrum f\"ur   
        Materialwissenschaften, Philipps Universit\"at     
	Marburg, D-35032 Marburg, Germany} 
\affiliation{Elm\'eleti Fizika Tansz\'ek, Fizikai Int\'ezet,  
        Budapesti M\H uszaki \'es Gazdas\'agtudom\'anyi Egyetem,  
        H-1521 Budapest, Hungary} 
\author{Peter Thomas} 
\affiliation{Fachbereich Physik und Wissenschaftliches Zentrum f\"ur   
        Materialwissenschaften, Philipps Universit\"at     
	Marburg, D-35032 Marburg, Germany} 
\author{Torsten Meier} 
\affiliation{Fachbereich Physik und Wissenschaftliches Zentrum f\"ur   
        Materialwissenschaften, Philipps Universit\"at     
	Marburg, D-35032 Marburg, Germany} 
\author{Stephan W. Koch} 
\affiliation{Fachbereich Physik und Wissenschaftliches Zentrum f\"ur   
        Materialwissenschaften, Philipps Universit\"at     
	Marburg, D-35032 Marburg, Germany} 
\date{\today} 
\begin{abstract} 
A non--equilibrium occupation distribution relaxes towards the Fermi--Dirac  
distribution due to electron--electron scattering even in finite  
Fermi systems. The dynamic evolution of this thermalization process assumed to
result from an optical excitation is investigated numerically by solving a 
Boltzmann equation for the carrier populations using a one--dimensional 
disordered system. We focus on the short time--scale behavior. The 
logarithmically long time--scale associated with the glassy behavior of 
interacting electrons in disordered systems is not treated in our investigation. 
For weak disorder and short range interaction we recover the expected result 
that disorder enhances the relaxation rate as compared to the case without 
disorder. For sufficiently strong disorder, however, we find an opposite 
trend due to the reduction of scattering probabilities originating from 
the strong localization of the single--particle states. Long--range interaction 
in this regime produces a similar effect. The relaxation rate is found to
scale with the interaction strength, however, the interplay between the 
implicit and the explicit character of the interaction produces an anomalous
exponent. 
\end{abstract}      
\maketitle{} 
 
The interplay of strong disorder and electron--electron interaction is one of 
the major issues of contemporary condensed matter physics. That problem may well be 
behind the insulator metal transition in two dimensions \cite{rev2dmit} or behind  
the unexpectedly large persistent current observed in experiments as  
compared to theoretical predictions \cite{persurr}. Similarly this interplay is
responsible for the glassy behavior of the electrons recently investigated both
experimentally \cite{Zvi} and theoretically \cite{VD}. Since theoretically  
the nonperturbative treatment of both disorder and interaction is still  
a very demanding task, numerical simulations may yield important insight
into the problem. In the present paper we present results of such a numerical
simulation.

For photoexcited ordered semiconductors it is 
known that Coulomb scattering is a rapid process \cite{stephan}. 
In the presence of weak disorder, i.e in dirty metals in the diffusive  
regime such process may become even faster \cite{boris} because the 
particles diffusively can spend more time close to each other, which 
results in an enhanced probability of scattering. This enhancement can 
in other words be attributed to the absence of $k$-vector selection
rules in the scattering process.  
 
Very little is known about the Coulomb scattering for the case of 
strong disorder. In this paper we show that with increasing disorder 
the localization length of the single--particle states reduces drastically 
and hence the scattering probabilities as well. Our result is obtained 
from numerical investigations of the dynamical energy  
relaxation due to electron--electron scattering in a system modeling 
a disordered metal in the localized regime. Our model is related to the  
quantum Coulomb--glass model introduced and studied in detail in  
Refs. [\onlinecite{qcglass}]. In those and subsequent studies that
model has been used to determine stationary and equilibrium properties 
of interacting electrons in a disordered environment. The interplay of
strong interaction with strong disorder is also responsible for the
emergence of glassy behavior \cite{Zvi,VD} which results in logarithmically
slow relaxation processes with time scales of the order of a day.

\begin{figure}[t]
\includegraphics[width=3.3in]{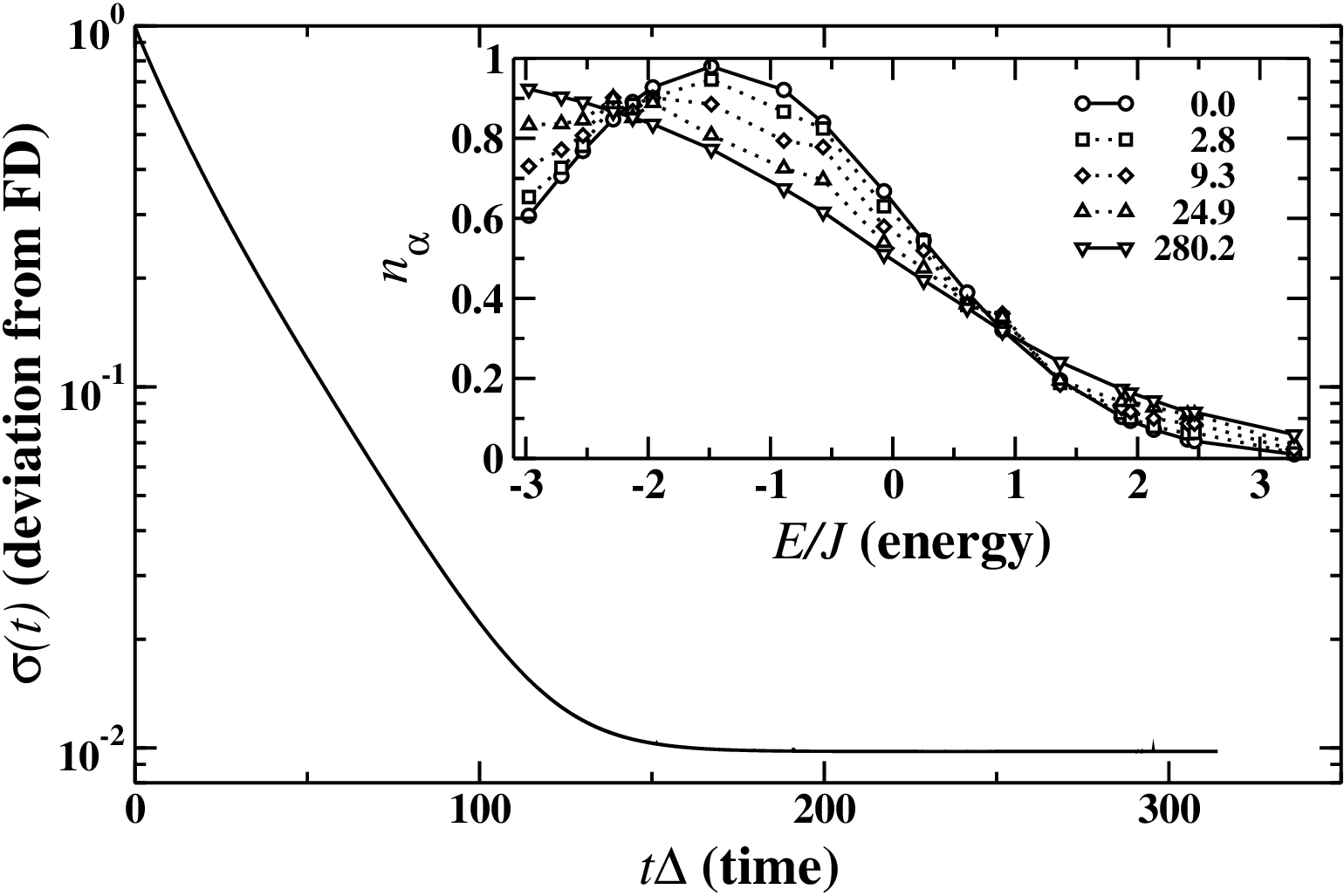} 
\caption{\label{fig:hiba}Typical time evolution of the  
root--mean--squared  
deviation of an initially non--equilibrium occupation from the 
Fermi--Dirac distribution. For the case of long--range Coulomb 
interaction. The inset shows the $n_{\alpha}$ distribution at  
different time instants. Time is measured in units of  
Heisenberg--time, $\Delta^{-1}$, energy in units of nearest 
neighbor hopping, $J$.  
} 
\end{figure} 

Here we are interested in the short time--scale relaxation of an initially 
non--equilibrium occupation number distribution which is assumed to be
a result of an optical excitation. Ultimately we are interested in 
optical phase relaxation due to Coulomb interactions in a strongly
disordered system. These particular relaxation processes will take place 
around a local minimum of the free energy of the Coulomb--glass. They are 
completed long before the system moves from one minimum to a lower one. 
Therefore the logarithmic relaxation times arising due to the slow
process of finding the global minimum of the free energy is out of the
scope of the present study. The treatment of optical relaxation due to
interactions is not a trivial subject even in ordered semiconductors
\cite{stephan}. In order to gain some insight into these processes we
here study the first the population relaxation in this time regime, 
typical for processes within a given minimum. We are aware, however, that
phase and population relaxations are not identical. Nevertheless, the
dependence of the relaxation rates on disorder and interaction strength in
the situation envisaged is interesting in itself as far as optical phenomena
are considered.

In order to investigate  
the diffusive and the localized regimes as well, we use both  
Hubbard--type short--range interactions and Coulomb--type  
long--range interactions. We already anticipate that the former 
is more appropriate in the diffusive regime as it roughly incorporates 
the screening effect of the other electrons, although, both types of 
interaction yield qualitatively similar results in the localized 
regime.
 
\begin{figure}[ht] 
\includegraphics[width=3.3in]{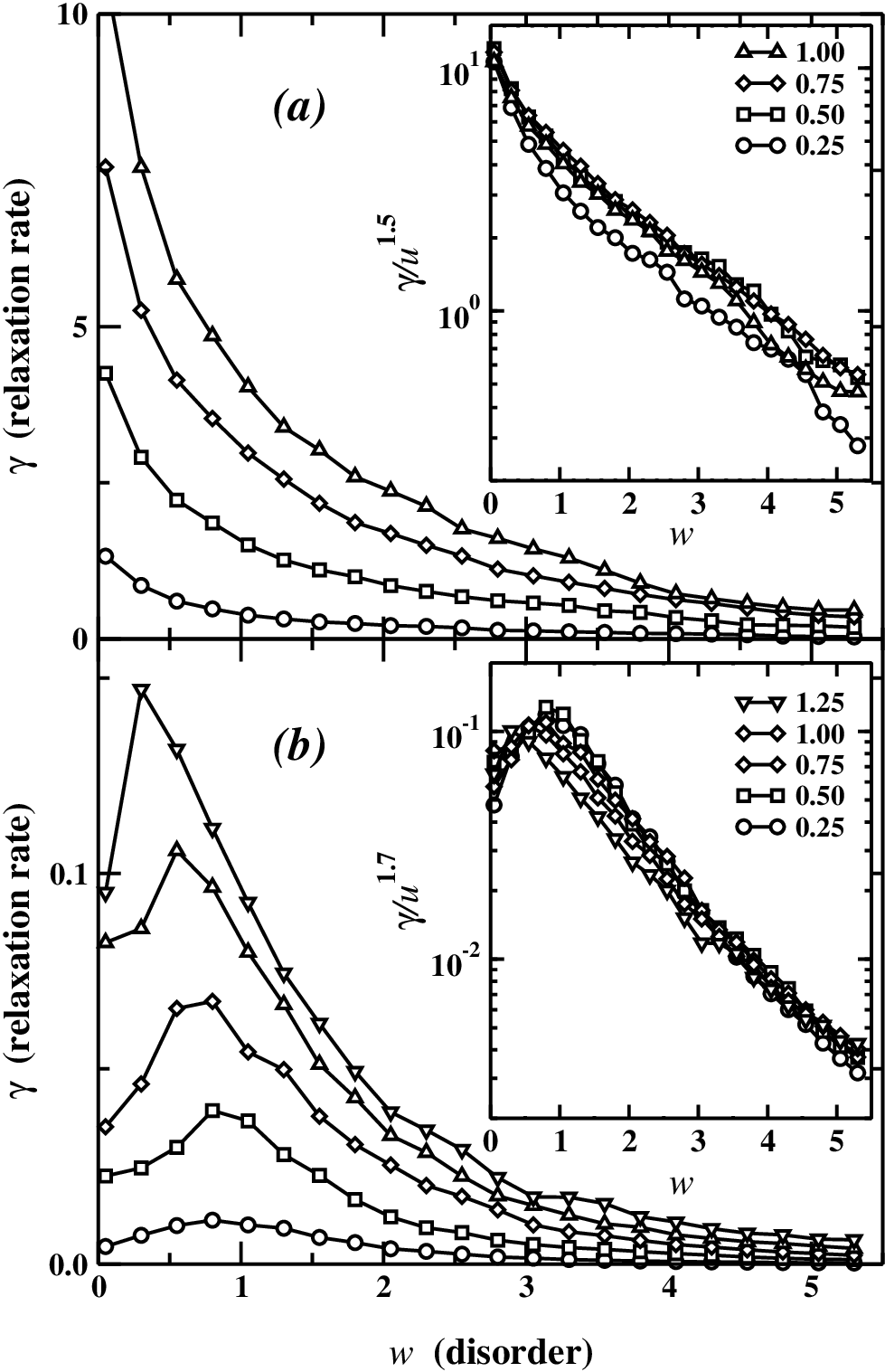} 
\caption{\label{fig:relt}Dimensionless relaxation rate in units of the  
mean level spacing for the case, $\gamma=\Gamma/\Delta$, of (a)  
long--range Coulomb interaction and (b) short--range Hubbard  
interaction as a function of the dimensionless disorder strength, 
$w=W/J$. The initial occupation is similar as in the inset of  
Fig.~\protect\ref{fig:hiba}. The different curves are labeled  
according to $u=U_0/J$. The insets show the curves rescaled with the 
dimensionless interaction parameter, $u$. The system size is $N=20$. 
} 
\end{figure}  
In order to investigate the population relaxation due to the 
particle--particle scattering we consider a simplified model of a
strongly disordered system described by the Hamiltonian that consists  
of two parts, $H=H_1 + H_2$, where the single--particle part, $H_1$ reads as 
\beq 
\label{h1}
H_1=\sum_i\varepsilon_i c^{\dag}_ic_i+ 
    \sum_{i,j}J_{ij}c^{\dag}_ic_j 
\eeq 
where $c_i$ ($c^{\dag}_i$) annihilates (creates) an electron on site  
$i$, (i.e. a state $|i\rangle$). We consider electrons without spin. 
The atomic energy levels, $\varepsilon_i$, are taken randomly from  
a box distribution of width $W$ around zero mean value. 
$J_{ij}$ describe the hopping amplitude from site $i$  
to site $j$. Nearest neighbor approximation has been used, with a constant
hopping rate $J$ taken as the unit of energy. The sites are  
assembled in a regular one--dimensional lattice of unit lattice spacing 
with periodic boundary conditions. 
 
The second part of the total Hamiltonian contains the 
two--particle interaction which in site representation reads as 
\beq 
H_2=\frac{1}{2}\sum_{ij}V_{ij}(c^{\dag}_ic_i-K)(c^{\dag}_jc_j-K), 
\eeq 
where for the sake of charge neutrality we have already included 
a compensating charge of $Ke$ at each lattice site where $K$ is the 
filling factor.  The interaction  
matrix element is either of short--range or long--range type.
In the former case $V_{ij}=U_0$ 
when two electrons are on the neighboring sites, $|i-j|=1$, and zero  
otherwise. 
For long--range interaction, 
$V_{ij}=U_0/|i-j|$, $U_0>0$ characterizes the strength of  
the repulsion between electrons located at neighboring sites.  
In any case due to the Pauli--principle
the electrons are not allowed to occupy the same site. 
 
The electron--electron scattering is evaluated in an effective 
single--particle basis.  
This basis is obtained from the diagonalization of the Hamiltonian $H_1$  
including the `diagonal' part of the interaction, i.e. as a first  
step we selfconsistently
obtain the Hartree--Fock (HF) solution of the Hamiltonian (\ref{h1})
by replacing the parameters $\varepsilon_i$ and $J_{ij}$ as
$\varepsilon_i+\frac{1}{2}\sum_jV_{ij}n_{jj}$ and 
$J\delta_{\langle i,j\rangle}-\frac{1}{2}V_{ij}n_{ji}$.
In the HF basis our original Hamiltonian can be expressed as 
\beq 
H=\sum_{\alpha}\varepsilon_{\alpha}c^{\dag}_{\alpha}c_{\alpha}+
  \sum_{\alpha\beta\gamma\delta}U_{\alpha\beta}^{\gamma\delta} 
   c^{\dag}_{\alpha}c^{\dag}_{\beta}c_{\gamma}c_{\delta} 
\eeq 
where the single--particle part, is obviously diagonal in the HF basis 
with the $\varepsilon_{\alpha}$'s being the HF eigenvalues. In the 
residual interaction 
\beq 
U_{\alpha\beta}^{\gamma\delta}=\sum_{ij}V_{ij} 
           C^*_{i\alpha}C^*_{j\beta}C_{i\gamma}C_{j\delta},
\label{intabcd} 
\eeq 
where the $\{C_{i\alpha}\}$ numbers represent the HF states in site
representation.
This model is in fact the standard two--body model of interacting (TBRIM)
fermions. Within the TBRIM, for example, the  
$U_{\alpha\beta}^{\gamma\delta}$ values are chosen randomly from a 
Gaussian distribution assuming that the single--particle states are 
sufficiently chaotic, i.e. delocalized. The typical matrix element 
$\bar{U}\approx \Delta/g$, where $g$ is the dimensionless conductance  
of the system and $\Delta$ is the mean HF--level spacing. 
In our case these interaction matrix elements contain the information  
about the microscopic details of the original model, e.g. the  
long--range correlations due to the Coulomb--interaction and the 
presence of a disorder potential. Hence, especially in the case of 
strong disorder we are not allowed to use the TBRIM model. Note also
that (\ref{intabcd}) depends on the interaction strength $U_0$ explicitly
through $V_{ij}$ and implicitly through the coefficients, $\{C_{i\alpha}\}$.
 
This HF basis corresponds to a zero temperature equilibrium  
distribution of the occupation numbers  
$n_{\alpha}=\langle c^{\dag}_{\alpha}c_{\alpha}\rangle$ that equals 
1 (0) for $\varepsilon_{\alpha}\leq E_F$ ($\varepsilon_{\alpha}>E_F$). 
 
We assume that an excitation process has somehow generated an initial,
non--equilibrium $n_{\alpha}$ distribution of the form 
$n_{\alpha}=Z^{-1}\exp\left[-(\varepsilon_{\alpha}-E_c)^2/2w^2\right]$,
where $E_c$ is the center of the `excitation' and $w$ is its energy spread. 
$Z$ is fixed by the condition $\sum_{\alpha}n_{\alpha}=N$. The center 
of the `excitation' is chosen in the lower half of the energy band, 
typically at its middle or at the bandedge. The width, $w$, is 
typically chosen to be one fourth of the bandwidth with which one could 
more-or-less avoid the possibility of the non-physical situation of 
$n_{\alpha}>1$. During the 
numerical simulation this initial distribution is assumed to relax towards
equilibrium via electron--electron scattering. This process  
is described by the Boltzmann equation \cite{boris} 
\begin{widetext} 
\bea 
\frac{d}{dt}n_{\alpha}=-\frac{2\pi}{\hbar}\sum_{\beta,\gamma,\delta} 
|U_{\alpha\beta}^{\gamma\delta}|^2\delta(\varepsilon_{\alpha}+ 
\varepsilon_{\beta}-\varepsilon_{\gamma}-\varepsilon_{\delta}) 
[n_{\alpha}n_{\beta}(1-n_{\gamma})(1-n_{\delta}) 
             -(1-n_{\alpha})(1-n_{\beta})n_{\gamma}n_{\delta}] 
\label{Born2} 
\eea 
\end{widetext} 
As we have noted already, in the localized regime we cannot apply  
assumptions of ergodic wave functions in order to estimate the typical  
value of $U_{\alpha\beta}^{\gamma\delta}$. Also our spectra are discrete 
therefore the $\delta$--function in (\ref{Born2}) is not possible to be
satisfied exactly. However, we may approximate it with a box of finite  
width of the order of the mean level spacing, $\Delta$. This approximation 
enables us to call the relaxation described by Eq.~(\ref{Born2}) in fact 
an inelastic process because the finite width effectively results in loss
of energy.
 
Integration of the above equation using a standard fourth order 
Runge--Kutta procedure gives the time evolution of $n_{\alpha}(t)$. 
The form of a Fermi--Dirac distribution 
$n_{FD}(E)=1/(1+\exp[\beta (E-\mu)])$
is fitted at every time step. This fit provides a `chemical potential', 
$\mu$ and an `inverse temperature', $\beta$. However, we are more  
interested in the error of this fit, 
$
\sigma^2(t)=\sum_{\alpha} 
       (n_{\alpha}-n_{FD}(\varepsilon_{\alpha}))^2. 
$
This quantity characterizes how close the distribution $n_{\alpha}$  
is to $n_{FD}$. An example for one single realization over $N=20$  
sites with half filling is presented in Fig.~\ref{fig:hiba} where we  
can clearly see an approximately exponential decrease that initially  
characterizes the relaxation process. From these initial exponentials 
a relaxation rate, $\Gamma$, can be obtained via 
${\sigma(t)}\approx \sigma_0\,\exp(-\Gamma t)$. 

The scattering probability between the pairs of 
single--particle states $\{\alpha,\beta\}$ and $\{\gamma,\delta\}$ is 
$(U_{\alpha\beta}^{\gamma\delta})^2$ which is explicitly proportional 
to $U_0^2$. Therefore for small enough $U_0$ when the Hartree--Fock  
states differ very little from the non--interacting basis we expect 
$\Gamma\sim U_0^2$. The power of two should, however, be an 
approximate value, since for strong enough interaction and also for 
strong enough disorder we expect a different exponent due to the implicit
character of the interaction encoded in the coefficients $\{C_{i\alpha}\}$.
 
The relaxation rate obtained from the exponentials as shown in 
Fig.~\ref{fig:hiba} has been collected and averaged over many  
realizations. In Fig.~\ref{fig:relt} we show data obtained  
for several interaction strength, $U_0$, and disorder, $W$. We can 
clearly identify qualitatively that for the case of short--range 
interactions, weak disorder produces an increase of the relaxation rate  
\cite{boris}. As disorder is increased, however, the rate 
decreases. Note that a long--range interaction induces  
a much faster relaxation as compared to a short--range one. 
 
As we can see in the insets of Fig.~\ref{fig:relt}  
the relaxation rate grows as a power of $U_0$ that is smaller than two. 
This is due to the strong perturbation the interaction makes on the HF 
states as compared to the non--interacting basis. One may also detect a
slight difference in the exponents between the two types of interaction.
 
\begin{figure}[t] 
\includegraphics[width=3.3in]{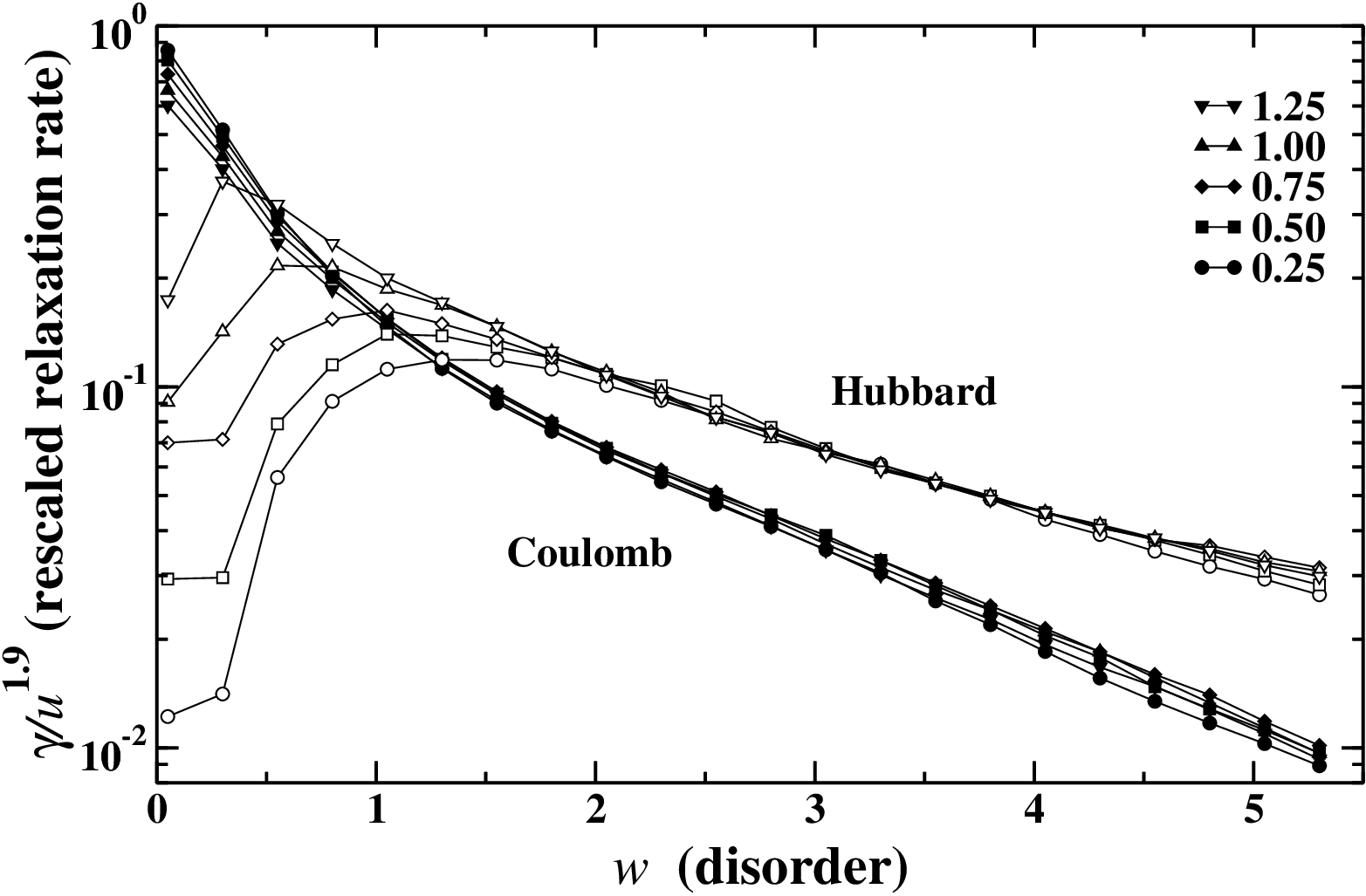} 
\caption{\label{fig:gammas}Average dimensionless relaxation rate,  
$\gamma=\langle\Gamma_{\alpha}\rangle/\Delta$ as a function of the  
dimensionless disorder strength, $w=W/J$ as obtained from the exponential
\textit{ansatz} for short times.
The different curves are labeled according to $u=U_0/J$. Open symbols 
stand for Hubbard type interaction and filled symbols for Coulomb type 
interaction. The data for the Coulomb interaction has been 
scaled down by a factor of $N$, where $N=20$ is the systems size. 
} 
\end{figure}

We may summarize that the error of the occupation number distribution 
decays for short times roughly as an exponential (Fig.~\ref{fig:hiba}) 
therefore we may expect that the solution of Eq.~(\ref{Born2}) is also 
an exponential whose derivative is $\dot{n}_{\alpha}\approx 
-\Gamma_{\alpha} n_{\alpha}$. 
If we substitute this \textit{ansatz} into Eq.~(\ref{Born2}) we can 
calculate the $\Gamma_{\alpha}$ values from the initial derivatives. 
This allows obviously for a much better statistics, however, will 
still strongly depend on the initial $n_{\alpha}$ distribution. An 
average over the individual rates, $\Gamma_{\alpha}$, and obviously 
over many realizations of the disordered potential is presented in 
Fig.~\ref{fig:gammas}. We can observe that the effective $U_0$ 
dependence here is roughly two owing to the simple approximation of the
exponential \textit{ansatz} for short times.
 
In summary our numerical results show that weak disorder indeed 
causes the energy relaxation of an initially non-equilibrium occupation 
number distribution towards a Fermi--Dirac distribution to become 
faster for the case of short--range interactions. On the other hand 
with an increase of disorder the single--particle 
localization volume decreases hence the quasi--particles have smaller  
chance to effectively scatter and therefore the relaxation rate decreases  
considerably. Note that the long--range type interaction  
is more effective, therefore $\Gamma$ is orders of magnitude larger as  
compared to the short--range interaction. However, qualitatively both of  
them produce a similar tendency as a function of large enough disorder, $W$.
From the results presented here one could expect that coherent phenomena 
at elevated density of particles may be more robust with respect to dephasing  
processes in strongly disordered systems. Thus one could suspect that these
systems may serve as a testing ground of coherent phenomena even at elevated 
carrier densities, especially in the case when these phenomena do require 
the presence of strong disorder anyway. One such phenomenon is for instance 
the current echo \cite{echo}.

The loss of coherence, i.e. dephasing, following interband photoexcitation
has been widely studied in semiconductor systems, where the optically 
induced coherence is monitored in the time domain by ultrafast nonlinear 
optical techniques \cite{wir,weg}. Using many-body theory it has, however,  
been shown that the dephasing of the interband polarization due to the 
Coulomb many-body interaction cannot be estimated directly from 
the carrier--carrier scattering rates but is a rather complicated phenomenon.
So-called {\it in} scattering contributions largely compensate
the dephasing provided by the usual {\it out} scattering 
terms \cite{stephan,wir,weg,frank}. 
As a result the dependence of the dephasing rate 
of the interband polarization on the carrier density $n$
is rather weak, see Ref.~\cite{weg} where for two- and three-dimensional 
systems a dependence of $\propto n^{1/3}$ was found.
It would be very interesting to include coherent contributions in
our type of approach and to investigate to what extent
such results are altered in the presence of disorder.

We are indebted to B. Altshuler, F. Izrailev, and Ph. Jacquod for  
stimulating discussions. This work is supported by the Deutsche  
Forschungsgemeinschaft (DFG) through project No.~KO816/8-1, by the  
Max-Planck Research Prize of the Humboldt and Max-Planck Societies, 
by the Alexander von Humboldt Stiftung, 
by OTKA, Grant Nos. T032116, T034832, and T042981,
by the European Community's Human Potential
Programme under contract HPRN-CT-2000-00144, \textit{Nanoscale Dynamics, 
Coherence and Computation},
and by the Center for Optodynamics, Philipps University, Marburg,  
Germany. T.M. thanks the DFG for support via a Heisenberg fellowship. 
 
\end{document}